\documentclass[fleqn]{vch-book}
\usepackage{amsmath}
\usepackage{amssymb}
\usepackage{makeidx}
\usepackage{graphicx}
\usepackage{times}
\usepackage{tabularx}
\usepackage{booktabs}
\renewcommand \thechapter {}
\renewcommand \thesection {\arabic{section}}

\renewcommand \thefigure {\arabic{figure}}

\newtheorem{theorem}{Theorem}
\newtheorem{prop}{Proposition}

\newcommand{\qed}{\hfill $\Box$}
\newcommand{\NN}{\mathbb{N}}
\newcommand{\ZZ}{\mathbb{Z}}
\newcommand{\QQ}{\mathbb{Q}}
\newcommand{\RR}{\mathbb{R}}
\newcommand{\CC}{\mathbb{C}}
\newcommand{\bs}[1]{\boldsymbol{#1}}

\renewcommand{\theenumi}{(\roman{enumi})}

\begin{document}

\chapter{Which distributions of matter diffract? --- Some answers}   

\authorafterheading{M.~Baake$^{\rm\,1)}$, R.~V.~Moody$^{\rm\,2)}$,
  C.~Richard$^{\rm\,1)}$ and B.~Sing$^{\rm\,1)}$}
\affil{\setlength{\tabcolsep}{0ex}
\begin{tabular}{rlrl}
$^{\rm 1)}$ & Institut f\"{u}r Mathematik & \qquad $^{\rm 2)}$ & Department of
Mathematical Sciences\\ 
& Universit\"{a}t Greifswald && University of Alberta \\
& Jahnstr.~15a && Central Academic Building \\
& 17487 Greifswald && Edmonton, Alberta T6G 2G1\\
& Germany && Canada \\
\end{tabular}}

\section{Introduction}
 
Diffraction experiments are important to determine the structure of a solid,
even more so with the refined methods available today. Recent applications
include aperiodic systems as well as systems with disorder.

Kinematic diffraction can be described and understood in terms of Fourier
analysis. The diffraction image is related to the Fourier transform of the
autocorrelation or Patterson function of the scattering obstacle, e.g., the
electron density. The situation is well understood for (ideal)
crystals, which show a complete lattice of periods, even though the
corresponding inverse problem does not have a unique solution in general.

Beyond the periodic situation, firm results are sparse, and until recently,
one did not even know which general distribution of scatterers would result in
a pure point diffraction spectrum, i.e., in a diffraction image of pure Bragg
peaks only. This review revolves around this question~\cite{BT86}, and
summarises the present state of affairs, with special emphasis on
contributions obtained during the time of the DFG focus program on
quasicrystals. We concentrate on the few results here that have been
established \textsl{rigorously}, and indicate where further research is needed.

The article is organised as follows: In the first section, we state 
conditions (Theorem~\ref{thm:pp_almostper}) under which a set of scatterers
diffract, i.e., its diffraction spectrum consists of Bragg peaks only. Instead
of looking at very general set of scatterers, we specialise in those sets that
model physical structures (as weighted Dirac
combs). Theorem~\ref{thm:dirac_comb} then states criteria under which these
sets diffract. With this result, we explain in Sec.~\ref{sec:model_sets} how a
cut and project scheme naturally appears. So, we obtain the associated
internal space of model sets, like the Fibonacci or the rhombic Penrose
tilings, through the information given by the autocorrelation. Model sets (or
cut and project sets) are the most common models for aperiodic order
(quasicrystals). 

In Sec.~\ref{sec:lattice} we investigate lattice substitution
systems. We first explore the diffraction spectrum for scatterers
distributed (aperiodic or even stochastic) over a lattice
(Theorems~\ref{thm:lattice} and~\ref{thm:lattice_compl}), before turning to
distributions obtained by substitutions. For this special case, we are
again interested under which conditions they diffract
(Theorem~\ref{thm:LMS03}). As an application of a lattice substitution
system as well as a model set with non-Euclidean internal space, we calculate,
in Sec.~\ref{sec:paperfolding}, the diffraction spectrum of the paperfolding
sequence.

Thereafter, we leave the area of deterministic point sets and turn to systems
with disorder, i.e., to random tilings in Sec.~\ref{sec:disorder}. We
carefully introduce the notion of a random tiling, before we state results
about one dimensional binary random tilings in
Theorem~\ref{thm:1Dbinary_random} and about the two dimensional Ising lattice
gas in Theorem~\ref{thm:Ising_gas}. We end this section by the example of a
Fibonacci random tiling, where we are particularly interested in the role of
the internal space of quasiperiodic random tilings for their diffraction
spectrum (Theorems~\ref{thm:intdistr} and~\ref{thm:R02}). We conclude our
article with an outlook (Sec.~\ref{sec:outlook}) where we indicate future
directions in diffraction theory. 

\section{Mathematical diffraction theory}

The basic object of interest is a set of scatterers in a
Euclidean\footnote
{
Most results also apply to more general spaces, namely
$\sigma$-compact locally compact Abelian groups, with Lebesgue measure
replaced by Haar measure, etc., see~\cite{BM02}.
} 
space $\RR^{d}$, which we model by a \textit{translation bounded complex
measure}\footnote
{
Frequently, we make use of the one-to-one correspondence
between \textit{measures} and regular \textit{Borel measures} by the
Riesz-Markov representation theorem, where
a \textit{measure} is a continuous linear functional on the space of compactly
supported continuous functions on $\RR^{d}$, while a \textit{Borel measure} is
defined on the Borel sets of $\RR^{d}$. The \textit{convolution} of two
measures $\mu$, $\nu$ is defined as
$\mu\ast\nu(f) = \int_{\RR^{d}\times\RR^{d}} f(x+y) \, d\mu(x)\, d\nu(y)$ and
is well-defined if at least one of them has compact support.
A measure $\mu$ is \textit{translation bounded} if, for all compact
$K \subset \RR^{d}$, $\sup_{t\in\RR^{d}} |\mu|(t+K) \le C^{}_{K} < \infty$ for
some constant $C^{}_{K}$ which only depends on $K$. Here, $|\mu|$ denotes the
\textit{total variation measure} (which is positive) and $t+K = \{ t+x
\mathbin| x\in K\}$. See~\cite{BM02,BF75,RS80,LA90,Hof95,La00} and references
therein for details.
} 
$\omega$. It describes the distribution of matter in a mathematically adequate
way. To calculate the diffraction spectrum, we need
the \textit{autocorrelation measure}\index{autocorrelation!measure}
$\gamma^{}_{\omega}$ attached to $\omega$. The \textit{diffraction
  spectrum}\index{diffraction spectrum} is then
the Fourier transform $\hat{\gamma}^{}_{\omega}$ of the autocorrelation
measure. Here, $\hat{\gamma}^{}_{\omega}(E)$ is the total intensity scattered
into the volume element $E$, and thus describes the outcome of a diffraction
experiment, compare~\cite{Cow95}. It can uniquely be decomposed as
$\hat{\gamma}^{}_{\omega} = (\hat{\gamma}^{}_{\omega})^{}_{\textnormal{pp}} +
(\hat{\gamma}^{}_{\omega})^{}_{\textnormal{sc}} +
(\hat{\gamma}^{}_{\omega})^{}_{\textnormal{ac}}$ by the Lebesgue decomposition
theorem, see~\cite{RS80}. 
Here, $(\hat{\gamma}^{}_{\omega})^{}_{\textnormal{pp}}$ is a \textit{pure point
  measure}, which corresponds to the Bragg part of the diffraction spectrum,
$(\hat{\gamma}^{}_{\omega})^{}_{\textnormal{ac}}$ is \textit{absolutely
  continuous} and $(\hat{\gamma}^{}_{\omega})^{}_{\textnormal{sc}}$
\textit{singular continuous} with respect to Lebesgue measure. The pure
pointedness of the diffraction spectrum $\hat{\gamma}^{}_{\omega}$ of $\omega$
is in question, i.e., whether the Lebesgue decomposition reduces to
$\hat{\gamma}^{}_{\omega} = (\hat{\gamma}^{}_{\omega})^{}_{\textnormal{pp}}$
(then, the diffraction spectrum consists of Bragg peaks only). 

We first fix an averaging sequence $\mathcal{A} = \{ B^{}_{n} \mathbin| n \in
\NN \}$  of balls of radius $n$ around $0$, so we begin analysing the spectrum
by looking at finite pieces (balls) of the structure in question.  
We define $\tilde{\omega}(f) = \overline{\omega(\tilde{f})}$, 
where $\tilde{f}(x)=\overline{f(-x)}$, and set 
$\omega^{}_{n} = \omega|^{}_{B^{}_{n}}$ and $\tilde{\omega}^{}_{n} =
(\omega^{}_{n})\tilde{\;}\,$. Then, the measures
\begin{equation}\label{eq:autocorr_general}
\gamma^{(n)}_{\omega} =
\frac{\omega^{}_{n}\ast\tilde{\omega}^{}_{n}}{\operatorname{vol}(B^{}_{n})}  
\end{equation}  
are well-defined, since they are (volume averaged) convolutions of two measures
with compact support. The autocorrelation $\gamma^{}_{\omega}$ of $\omega$
exists, if $(\gamma^{(n)}_{\omega})^{}_{n\in \NN}$ converges in the vague
topology, compare~\cite{Hof95}, and is then, by construction, a positive
definite\footnote
{
A measure $\mu$ is \textit{positive definite} iff
$\mu(f\ast\tilde{f}) \ge 0$ for all complex-valued continuous functions with
compact support, compare~\cite{BF75}.
} 
measure. 

We say that a measure is \textit{almost periodic} if the set of translates is
\textit{relatively compact} (i.e., its closure is compact). Of course, we have
to fix a topology for this, and in our case we need the \textit{strong} or
\textit{product topology} on the space of translation bounded complex
measures (and not the vague topology), see~\cite{BM02,LA90}. With this
topology, we speak of \textit{strong almost periodicity} to distinguish it
from almost periodicity in other topologies. The key result reads:

\begin{theorem}\label{thm:pp_almostper}{\rm\cite{BM02,LA90}}
The measure $\omega$ is pure point diffractive iff
\begin{enumerate}
\item\label{thm:item1:pp_almostper} $\gamma^{}_{\omega}$ exists.
\item $\gamma^{}_{\omega}$ is strongly almost periodic.\qed
\end{enumerate}
\end{theorem}
Note that~\ref{thm:item1:pp_almostper} is merely a convention, since one can
always pick an appropriate subsequence of the averaging sequence $\mathcal{A}$
for which $\gamma^{}_{\omega}$ exists.
 
So, given a measure $\omega$, we have to check these two conditions to decide
whether it is pure point diffractive. This can be done for many relevant
examples. We would also like to solve the \textit{homometry} or
\textit{inverse problem}, i.e., the question which measures $\omega$ account
for a given diffraction spectrum $\hat{\gamma}$. This is a hard problem,
because there is no inversion process of Eq.~\ref{eq:autocorr_general}. In
fact, rather different $\omega$ can have the same diffraction,
see~\cite{HB00}. One also would like to understand the implications of a pure
point spectrum. Here, we will not explore these last two questions further.

Instead, let us specialise on the situation of a countable set
$S$ of scatterers in $\RR^{d}$ with (bounded) scattering strengths 
$v(x)$, $x \in S$. It can be represented as a complex Borel measure
in the form of a \textit{weighted Dirac comb}\footnote
{
For later reference, we write $\omega = A \cdot \delta^{}_{S}$ if $v(x) = A$
for all $x\in S$.
}
\begin{equation*}
\omega = \sum_{x \in S} v(x)\,\delta^{}_{x},
\end{equation*} 
where $\delta^{}_{x}$ is the unit point (or Dirac) measure located at
$x$, i.e., $\delta^{}_{x}(\varphi) = \varphi(x)$ for continuous functions
$\varphi$. This way, atoms are modeled by their position and scattering
strengths. Convolutions with more realistic profiles are not considered here,
but can easily be treated by the convolution theorem, see~\cite{BF75}.

Denote the set of ``inter-atomic distances'' by $\varDelta=S-S = \{x-y
\mathbin| x,y \in S\}$. Then we make the following three
assumptions, see~\cite{BM02}: 
{\newcounter{assume}
\begin{list}{(A \arabic{assume})\hfill}
{\usecounter{assume}
\addtolength{\itemindent}{-.64em}
\addtolength{\leftmargin}{.64em}
}
\item\label{as:1} The measure $\omega$ is translation bounded, i.e, 
there exist constants $C^{}_{K}$ so that
\begin{equation*}
\sup\limits_{t\in\RR^{d}} \sum_{x \in S \cap (t+K)} |v(x)| \; \le C^{}_{K} <
\infty 
\end{equation*}
for all compact sets $K \subset \RR^{d}$.
\item\label{as:2} The \textit{autocorrelation coefficients}
\begin{equation*}
\eta(z) = 
\lim\limits_{n \to \infty} \frac{1}{\operatorname{vol}(B^{}_{n})}
\sum\limits_{\substack{x,y \in S \cap B^{}_{n} \\ x-y = z}} v(x) \,
\overline{v(y)} 
\end{equation*}
exist for all $z \in \varDelta$ (we set $\eta(z) = 0$ if $z \not\in
\varDelta$). 
Consequently (if also (A \ref{as:3}) holds), the autocorrelation
measure $\gamma^{}_{\omega} = \sum_{z \in \varDelta} \eta(z) \, \delta^{}_{z}$
exists. 
\item\label{as:3} The set $\varDelta^{\textnormal{ess}}_{} = \{ z \in
  \varDelta \mathbin| \eta(z)\neq 0\} \subset \varDelta$ is \textit{uniformly
  discrete}, i.e., there is an $r>0$  such that open balls of radius $r$
  centred at the points of $\varDelta^{\textnormal{ess}}_{}$ are mutually
  disjoint.  
\end{list}}
The support of $\gamma^{}_{\omega}$ (which is the support of
$\eta(z)$) plays a special role, in particular the group $L$ generated by it:
\begin{equation}\label{eq:defL}
L = \left< \varDelta^{\textnormal{ess}}_{} \right>^{}_{\ZZ} \subset \RR^{d}
\end{equation}

The point set $S$ is \textit{repetitive} if for any compact set 
$K \subset \RR^d$, $\{t \in \RR^d \mathbin| S \cap K = (t + S) \cap K\}$ is
relatively dense\footnote
{ 
A set $Q$ is \textit{relatively dense} if there is a radius $R>0$
such that every ball of radius $R$ in $\RR^{d}$ contains at least one point of
$Q$.
}; 
i.e., there exists a radius $R = R(K) > 0$ such that every open ball 
$B_R(y)$ contains at least one element of $t \in \RR^d $ for which $ S \cap K
= (t + S) \cap K $. If $S$ is repetitive, we have $\varDelta =
\varDelta_{}^{\textnormal{ess}}$, hence $L = \left< \varDelta
\right>^{}_{\ZZ}$. 

We define a \textit{translation invariant pseudo-metric}\footnote
{
A \textit{pseudo-metric} is a non-negative, symmetric function on $\RR^{d}
\times \RR^{d}$ that satisfies the triangle inequality. 
Such a function $\varrho$ is \textit{translation invariant} if 
$\varrho(s+r,t+r) = \varrho(s,t)$ for all $r\in \RR^{d}$.
} 
by  
\begin{equation}\label{eq:pseudometric}
\varrho(s,t) = \left| 1 - \frac{\eta(s-t)}{\eta(0)} \right|^{\frac12}_{}.
\end{equation}
If all weights $v(x)$ are non-negative, one has $\varrho(s,t) \le 1$ for all
$s,t \in \RR^{d}$, but, in general, $\varrho$ is bounded by $\sqrt{2}$.
This pseudo-metric $\varrho$ defines a \textit{uniformity},
both on $L$ and $\RR^{d}$, see~\cite{BM02}. The induced topology
is, in general, completely different from the usual Euclidean topology of
$\RR^{d}$. It is called the \textit{autocorrelation
  topology}\index{autocorrelation!topology}. 

Next we define, for $\varepsilon > 0$, a set $P^{}_{\varepsilon}$ of
$\varepsilon$\textit{-almost periods} of the autocorrelation
$\gamma^{}_{\omega}$ through
\begin{equation}\label{eq:almostper}
P^{}_{\varepsilon} = \{ t \in \RR^{d} \mathbin| \varrho(t,0) < \varepsilon \}.
\end{equation}
We clearly have the following inclusions for $\varepsilon < \varepsilon'$:
\begin{equation*}
\{\mbox{periods of }\omega\} \quad \subset \quad P^{}_{\varepsilon} \quad
\subset \quad P^{}_{\varepsilon'} \quad \subset \quad \RR^{d},
\end{equation*}
and furthermore $P^{}_{1} = \varDelta^{\textnormal{ess}}_{}$ if all weights
$v(x)$ are non-negative. Now, we are able to apply
Theorem~\ref{thm:pp_almostper} to weighted Dirac combs.

\begin{theorem}\label{thm:dirac_comb}{\rm\cite{BM02}}
Let $\omega$ be a weighted Dirac comb that satisfies \textnormal{(A
  \ref{as:1})}, \textnormal{(A \ref{as:2})} and \textnormal{(A
  \ref{as:3})}. Then $\hat{\gamma}^{}_{\omega}$ exists, and the following
statements are equivalent: 
\begin{enumerate}
\item $P^{}_{\varepsilon}$ is relatively dense for all $\varepsilon > 0$.
\item\label{thm:item2:dirac_comb}$\gamma^{}_{\omega}$ is norm almost periodic.
\item $\gamma^{}_{\omega}$ is strongly almost periodic.
\item $\hat{\gamma}^{}_{\omega}$ is pure point
  diffractive. \qed 
\end{enumerate}
\end{theorem}
The \textit{norm almost periodicity} in~\ref{thm:item2:dirac_comb} refers to
the topology defined by the norm $\|\omega\|^{}_{K} = \sup_{x\in\RR^{d}}
|\omega|(x+K)$ for some fixed compact $K$ with nonempty interior, e.g., a
closed unit ball. The concept of norm almost periodicity is independent of the
choice of $K$. 

This theorem applies to the diffraction from the visible lattice points,
see~\cite{BM02,BMP99}, but also to model sets which we will consider next.

\section{Model sets\label{sec:model_sets}}

The well-known cut and project mechanism provides examples of weighted point
sets which satisfy the assumptions (A~\ref{as:1}), (A~\ref{as:2}) and
(A~\ref{as:3}). But here, we start with a countable weighted set of
scatterers in $\RR^{d}$ for which the corresponding weighted Dirac comb
fulfils Theorem~\ref{thm:dirac_comb} and is therefore pure point
diffractive. This will lead us to a suitable cut and project scheme.

The constructive picture behind this is the following: On $L$
of Eq.~\ref{eq:defL}, we have a pseudo-metric $\varrho$, defined
in Eq.~\ref{eq:pseudometric}, which in turn defines a uniformity, and then
the \emph{autocorrelation topology}\index{autocorrelation!topology} on
$L$. Inside $L$, the sets $P^{}_{\varepsilon}$ of Eq.~\ref{eq:almostper} are
the open balls of radius $\varepsilon$ around $0$ in the autocorrelation
topology. The group $L$ equipped with this topology admits a
\textit{(Hausdorff) completion}\footnote 
{
$H$ is a \textit{completion} of $L$, if $L$ has dense image in $H$ and every
Cauchy sequence in $H$ has a limit in $H$, e.g., $\RR$ is the completion of
$\mathbb{Q}$. The completion is unique up to topological isomorphism.
} 
$H$, which is a \textit{locally compact Abelian group}\footnote
{
A topological space is called \textit{locally compact} if each point is
contained in a compact neighbourhood. If this space is an Abelian group, we
speak of a locally compact Abelian group. Examples are Euclidean and
$p$-adic spaces.
}. 
This means that there is a continuous group homomorphism
$\varphi\!\!: L \to H$ where $\varphi(L)$ is dense in $H$. 
The relative denseness of $P^{}_{\varepsilon}$ is crucial here, because for
every set $P^{}_{\varepsilon}$ of $L$ there is an open set $B(\varepsilon)$ of
$H$ so that $\varphi(P^{}_{\varepsilon}) = \varphi(L) \cap B(\varepsilon)$ and
$B(\varepsilon)$ has compact closure $\overline{B(\varepsilon)}$ for
$0<\varepsilon < 1$, giving the local compactness of $H$. 

We obtain the following \textit{cut and project scheme}\index{cut and project
  scheme}, see~\cite{BM02}: 
\begin{equation}\label{eq:cut_and_project}
\renewcommand{\arraystretch}{1.5}
\begin{array}{ccccc}
& \pi & & \pi_{\textnormal{int}}^{} & \vspace*{-1.5ex} \\
\RR^{d} & \longleftarrow & \RR^{d}\times H & \longrightarrow &
H = \overline{\varphi(L)} \\ 
\cup  & \mbox{\raisebox{-1.5ex}{\footnotesize
     \textnormal{1--1}}}\!\!\!\!\nwarrow\;\; & \cup &  
\;\;\nearrow\!\!\!\!\mbox{\raisebox{-1.5ex}{\footnotesize \textnormal{dense}}}
& \\
L & \longleftrightarrow & \tilde{L}=\{(x,\varphi(x))\mathbin| x\in L\} & &
\end{array}
\end{equation} 
Here, $\tilde{L}$ is a \textit{lattice} in $\RR^{d} \times H$, i.e.,
$\tilde{L}$ is a closed subgroup so that the
factor group $(\RR^{d} \times H)/\tilde{L}$ is compact. The projection
$\pi_{\textnormal{int}}$ is dense in \textit{internal space}\index{internal
  space} $H$ and the projection $\pi$ into \textit{physical
  space}\index{physical space} $\RR^{d}$ is one-to-one on $\tilde{L}$.  

The internal space $H$ and the lattice $\tilde{L}$ both arise from the group
$L$ via the autocorrelation topology. In spite of its abstraction, the
procedure gives back the familiar Euclidean internal spaces of the well-known
examples (e.g.\ Fibonacci). In the case of the rhombic Penrose tilings 
it gives the minimal internal space possible for representing them in the cut
and project formalism: $H= \RR^{2} \times (\ZZ/5 \ZZ)$, see~\cite{BM02}. 
For other tilings, the internal space may be $p$-adic, see
Section~\ref{sec:paperfolding} for an example. 
Note that the completion map $\varphi$ is not one-to-one in general, its
kernel is $\bigcap_{\varepsilon>0} P^{}_{\varepsilon}$, the group of
statistical periods. For example, if $S$ is a lattice with all weights equal,
then $S=L$, $\varphi \equiv 0$, $H=0$ and $\tilde{L} \cong L$, so the cut and 
project scheme collapses into triviality. In general, the internal space $H$ 
ignores the periodic part of $\omega$, for which no additional structure is 
required, and reflects only the aperiodic parts.

A set $\varLambda \subset \RR^{d}$ is a \textit{model set}\index{model set}
for the cut and project scheme in Eq.~\ref{eq:cut_and_project}, if there is a
relatively compact set $W \subset H$ with non-empty interior and a $t\in\RR^d$
such that 
\begin{equation*}
\varLambda = t+\varLambda(W) = t+\{ x \in L \mathbin| \varphi(x) \in W \}.
\end{equation*} 
Note that, in the context of model sets, the map $\varphi$ is often called the
$\star$-map and denoted by $(\cdot)_{}^{\star}$, i.e., one writes
$x_{}^{\star} = \varphi(x)$. 
A model set is always a \textit{Delone set}\index{Delone!set}, i.e., it is
both relatively dense and uniformly discrete. A model set is called
\textit{regular} if the boundary of $W$ has \textit{Haar measure}\footnote
{
On every locally compact Abelian group there exists a unique 
(up to a multiplicative constant) translation invariant regular Borel measure. 
This is called the Haar measure, and is given by the Lebesgue measure on 
Euclidean space $\RR^{d}$.
} 
$0$. Regular model sets are the most relevant model sets for the physical
applications in the theory of quasicrystals. They also play a prominent role
in the analysis of sequences with long-range (aperiodic) order,
cf.~\cite{BS02} and references therein. 
 
One of the cornerstones of the theory of model sets is:

\begin{theorem}{\rm\cite{Hof95, Sch00}} 
Regular model sets are pure point diffractive. \qed
\end{theorem}

Let us now go back to our discussion of diffraction in the context of
the assumptions (A~\ref{as:1}), (A~\ref{as:2}) and (A~\ref{as:3}). The pure
point diffraction in Theorem~\ref{thm:dirac_comb} is intimately
related to the cut and project scheme in Eq.~\ref{eq:cut_and_project}. But it
can happen that  the set $S$ itself is not a model set (e.g., as in the case of
the visible lattice points), see~\cite{BM02}. So the question arises: Which
pure point diffractive point sets \textsl{are} (regular) model sets? We have
only partial progress on this. 

Noting that $W^{}_{\varepsilon}=\overline{\varphi(P^{}_{\varepsilon})}$ has
non-empty interior for all $0< \varepsilon < 1$, it follows form our above
discussion that $P^{}_{\varepsilon} \subset
\varLambda(W^{}_{\varepsilon})$. Since $L$ is countable, we even get that
$P^{}_{\varepsilon} = \varLambda(W^{}_{\varepsilon})$ can be violated
for at most countably many values of $\varepsilon$, so that
$\varDelta_{}^{\textnormal{ess}}$ is the union of an ascending sequence of
model sets. Furthermore we have

\begin{prop}{\rm\cite{BM02c}} 
Assume \textnormal{(A~\ref{as:1})}, \textnormal{(A~\ref{as:2})} and
\textnormal{(A~\ref{as:3})} hold. Then $\varDelta_{}^{\textnormal{ess}}$ is
a model set.\qed 
\end{prop}

This leaves open the question of whether or not $\varLambda$ is a model set.
Progress in this seems to depend on utilising the dynamical hull of
$\varLambda$: 
\begin{equation}\label{eq:dynamical_system}
X(\varLambda) = \overline{\{\varLambda + t \mathbin| t \in \RR^{d}\}}, 
\end{equation}
where closure is taken with respect to the \textit{local topology}\footnote
{
Informally, two discrete and closed point sets are close in the local
topology, if, after a small translation, these two coincide on a large compact
region.
}. 
Then $\left(X(\varLambda),\RR^{d}\right)$ is a \textit{dynamical
  system}\index{dynamical system} under
the obvious action of $\RR^{d}$ on $X(\varLambda)$. 

If $\varLambda$ is a repetitive regular model set then it is known
\cite{Sch00} that $X(\varLambda)$ is \textit{strictly ergodic}, i.e., both
minimal\footnote 
{ 
A dynamical system $(X,T)$ is \textit{minimal} in case $X$ has no 
proper closed $T$-invariant subsets. It is
\textit{uniquely ergodic} if there is only one $T$-invariant Borel
probability measure on $X$.
}
and uniquely ergodic\footnotemark[\value{footnote}]. 
There is a conjecture to the effect that, conversely, any
pure point diffractive \textit{Meyer set}\index{Meyer set}\footnote 
{
A Delone set $S$ is a \textit{Meyer set} iff the set of ``inter-atomic
distances'' $\varDelta = S - S$ is also a Delone set. Every model set is a
Meyer set. 
}
of $\RR^{d}$ for which $X(\varLambda)$ is strictly ergodic is in fact a model
set. 

\section{Lattice substitution systems\label{sec:lattice}}

An interesting class of point sets is formed by the subsets of a lattice. Even
though they can be aperiodic or even stochastic, the underlying lattice
$\varGamma$ leaves its imprint, most notably in form of a periodicity of the
diffraction, with the dual lattice $\varGamma^{*}_{}$ as lattice of periods.

In a more general formulation, let $v\!\!: \varGamma \to \CC$ be any bounded
function, and consider the weighted Dirac comb
\begin{equation*}
\omega = \sum_{x \in \varGamma} v(x) \, \delta^{}_{x}.
\end{equation*}
This includes the previous case via $v = 1^{}_{S}$, i.e., $v(x) = 1$ for $x
\in S$ and $0$ otherwise. Then, if $\gamma^{}_{\omega}$ is any of its
autocorrelations (e.g.\ as obtained along a suitable subsequence of averaging
balls), we have the following result.

\begin{theorem}\label{thm:lattice}{\rm\cite{B00}}
Let $\varGamma$ be a lattice\footnote
{\newcounter{footnote:B00}\setcounter{footnote:B00}{\value{footnote}}
These results can be generalised to lattice subsets in locally compact Abelian
groups whose topology has a countable base, see~\cite{B00} for details.
} 
in $\RR^{d}_{}$ and $\omega$ a weighted Dirac comb on $\varGamma$ with bounded
complex weights. Let $\gamma^{}_{\omega}$ be any of its autocorrelations,
i.e., any of the limit points of the family $\{\gamma^{(n)}_{\omega} \mathbin|
n \in \NN\}$. Then the following holds. 
\begin{enumerate}
\item The autocorrelation $\gamma^{}_{\omega}$ can be represented as
\begin{equation*}
\gamma^{}_{\omega} = \varTheta \cdot \delta^{}_{\varGamma} = \sum_{x \in
  \varGamma} \varTheta(x) \, \delta^{}_{x},
\end{equation*}
where $\varTheta\!\!: \RR^{d}_{} \to \CC$ is a bounded continuous positive
definite function that interpolates the autocorrelation coefficients $\eta(x)$
as defined at $x \in \varGamma$. Moreover, there exists such a $\varTheta$
which extends to an entire function $\varTheta\!\!: \CC_{}^{d} \to \CC$ with
the additional growth restriction that there are constants $C, R \ge 0$ and $N
\in \ZZ$ such that $|\varTheta(z)| \le C \cdot ( 1+ |z|)^{N} \cdot \exp(R \:
|\operatorname{Im}(z)|)\:$ for all $z \in \CC_{}^{d}$.
\item The diffraction spectrum $\hat{\gamma}^{}_{\omega}$ of $\omega$ is a
  translation bounded positive measure 
  that is periodic with the dual lattice\footnote
{
The Euclidean scalar product is denoted by $\langle \cdot , \cdot \rangle$.
} 
$\varGamma^{*}_{} = \{k \in \RR^{d}_{} \mathbin| \langle k, x \rangle \in \ZZ
  \, \mbox{ for all } x \in \varGamma \}$ as lattice of periods. Furthermore,
  $\hat{\gamma}^{}_{\omega}$ has a representation as a convolution, 
\begin{equation*}
\hat{\gamma}^{}_{\omega} = \varrho \ast \delta^{}_{\varGamma^{*}_{}},
\end{equation*}
in which $\varrho$ is a finite positive measure supported on a fundamental
domain of $\varGamma^{*}_{}$ that is contained in the ball of radius $R$
around the origin. \qed
\end{enumerate}
\end{theorem}

This reduces the analysis of the spectral type of $\hat{\gamma}^{}_{\omega}$
to that of $\varrho$, which has compact support.

An interesting application concerns lattice subsets and their complements,
which leads to the following result.

\begin{theorem}\label{thm:lattice_compl}{\rm\cite{B00}}
Let $\varGamma$ be a lattice\footnotemark[\value{footnote:B00}]  
in $\RR^{d}_{}$, and let $S \subset \varGamma$ be a subset with existing
(natural) autocorrelation coefficients $\eta^{}_{S}(z) =
\operatorname{dens}\left(S\cap (z+S)\right)$. Then the following holds. 
\begin{enumerate}
\item The autocorrelation coefficients $\eta^{}_{S_{}^{c}}(z)$ of the
  complement set $S_{}^{c}= \varGamma\setminus S$ also exist. They are
  $\eta^{}_{S_{}^{c}}(z) = 0\:$ for all $z \not\in \varGamma$ and otherwise,
  for $z \in \varGamma$, satisfy the relation
\begin{equation*}
\eta^{}_{S_{}^{c}}(z)-\operatorname{dens}(S_{}^{c}) =
\eta^{}_{S}(z)-\operatorname{dens}(S) 
\end{equation*}
\item If, in addition, $\operatorname{dens}(S) =
  \operatorname{dens}(\varGamma)/2$, the sets $S$ and $S_{}^{c}$ are
  homometric. 
\item The diffraction spectra of the sets $S$ and $S_{}^{c}$ are related by
\begin{equation*}
\hat{\gamma}^{}_{S_{}^{c}} = \hat{\gamma}^{}_{S} + \left(
  \operatorname{dens}(S_{}^{c}) - \operatorname{dens}(S)\right)\cdot
  \operatorname{dens}(\varGamma) \, \delta^{}_{\varGamma^{*}_{}}.
\end{equation*} 
In particular, $\:\hat{\gamma}^{}_{S_{}^{c}} = \hat{\gamma}^{}_{S}\:$ if
$\:\operatorname{dens}(S_{}^{c}) = \operatorname{dens}(S)$. 
\item The diffraction measure $\hat{\gamma}^{}_{S_{}^{c}}$ is pure point
  iff $\hat{\gamma}^{}_{S}$ is pure point. \qed
\end{enumerate}
\end{theorem}

As an immediate consequence of part (i), one can check that the two
pseudometrics defined by $S$ and $S_{}^{c}$ via Eq.~\ref{eq:pseudometric}
are scalar multiples of one another, hence define the same uniformity
(as long as  
$\operatorname{dens}(S) \cdot \operatorname{dens}(S_{}^{c})>0$).

Of considerable interest in this context are lattice subsets which are
obtained by lattice substitution systems via \textit{Delone
  multisets}\index{Delone!multiset}\footnote 
{
A \textit{multiset} in $\RR_{}^{d}$ is a subset $U^{}_{1} \times \cdots \times
U^{}_{m} \subset \RR^{d}_{} \times \cdots \times \RR^{d}_{}$ ($m$ copies),
where $U^{}_{i} \subset \RR^{d}_{}$. We also write $\bs{U} = (U^{}_{1}, \ldots,
U^{}_{m})_{}^{t} = (U^{}_{i})^{}_{i\le m}$. We say that $(U^{}_{i})^{}_{i\le
  m}$ is a \textit{Delone multiset} in $\RR^{d}_{}$ if each $U^{}_{i}$ is
Delone and the union $\bigcup_{i=1}^{m} U^{}_{i} \subset \RR^{d}_{}$ is also
Delone. It is convenient to think of a multiset as a set with types of colours
(types of atoms), $i$ being the colour of points in $U^{}_{i}$.
}. 
They are the natural generalisation of one dimensional substitution rules with
constant length, and it was recently possible \cite{LM01,LMS02, LMS03} 
to find a complete generalisation of Dekking's coincidence criterion
(see~\cite{Dek78}) to this general situation. The result, stated below,  is a
circle of equivalences which directly puts pure pointedness and model sets  on
the same footing. Although one of the equivalent criteria is modular
coincidence, we refer the reader to~\cite{LMS03} for the rather technical
definition, just pointing out that its primary virtue is that it is  testable
by a straightforward algorithm. 

A \textit{matrix function system}\index{matrix function
  system}\index{MFS|see{matrix function system}} (MFS) on a lattice $\varGamma
\subset \RR^{d}_{}$ is given by an $m\times m$-matrix $\varPhi =
(\varPhi^{}_{ij})$, where each $\varPhi^{}_{ij}$ is a finite set (possibly
empty) of mappings $\varGamma \to \varGamma$. Here, the mappings of $\varPhi$
are affine linear mappings, where the linear part has the form $x \mapsto Q \,
x$ and is the \textsl{same} for all maps. We call $Q$ the \textit{inflation
  factor}. It is required to have all eigenvalues exceeding $1$ in absolute
value. Any MFS $\varPhi$ induces a mapping or \textit{substitution} on
$P(\varGamma)_{}^{m}$, where $P(\varGamma)$ denotes the set of subsets of
$\varGamma$: 
\begin{equation}\label{eq:MFS}
\varPhi \left( \begin{array}{c} U^{}_{1} \\ \vdots \\ U^{}_{m} \end{array}
\right) = \left( \begin{array}{c} \bigcup\limits_{j=1}^{m} \, \bigcup\limits_{f
      \in \varPhi^{}_{1j}} f(U^{}_{j}) \\ \vdots \\ \bigcup\limits_{j=1}^{m} \,
      \bigcup\limits_{f \in \varPhi^{}_{mj}} f(U^{}_{j}) \end{array} \right)
\end{equation}

We say that $\bs{U} = (U^{}_{1}, \ldots , U^{}_{m})_{}^{t}$ 
is a \textit{fixed point} of $\varPhi$ if $\bs{U} = \varPhi
\bs{U}$. Furthermore, we call $(\bs{U}, \varPhi)$ a \textit{lattice
  substitution system} on $\varGamma$ if $\varPhi$ is an MFS on $\varGamma$,
$\bs{U}$ is a fixed point of $\varPhi$, the $U^{}_{i}$'s are pairwise disjoint
and all the unions in Eq.~\ref{eq:MFS} are disjoint. A lattice substitution
system is \textit{primitive} if its corresponding \textit{substitution matrix}
$M = (\#\varPhi^{}_{ij})^{}_{1\le i,j \le m}$ is primitive, i.e., if there is
a $k \in \NN$ such that $M^{k}_{}$ has positive entries only.   

Let $\bs{U}$ be a lattice substitution system for the MFS $\varPhi$.
We say that a $\bs{U}$-cluster $\bs{C} = \bs{U} \cap B^{}_R(s) = (U^{}_{i}
\cap B^{}_R(s))^{}_{i \le m}$, defined by
intersecting all the components of $\bs{U}$ with a ball of radius $R$ around a
point $s\in \RR^d$, is \textit{legal} if it lies in $\varPhi_{}^n(\bs{u})$ for
some point $\bs{u}$ of $\bs{U}$. Furthermore, we define the symmetric
difference of two multisets as symmetric difference of their corresponding
components, i.e., $\bs{U}\triangle\bs{V} = (U^{}_{i}\triangle V^{}_{i})^{}_{i
  \le m} = ( (U^{}_{i}\setminus V^{}_{i}) \cup (V^{}_{i}\setminus
U^{}_{i}))^{}_{i \le m}$. We also use the notation $S + T = \{ x+y \mathbin| x
\in S, \, y \in T \}$  and $\bs{U} + S = (U^{}_{i} + S)^{}_{i \le
  m} = (\{ x+y \mathbin| x \in U^{}_{i}, \, y \in S \} )^{}_{i \le m}$ for
sets $S, T$ and a multiset $\bs{U}$.

\begin{theorem}\label{thm:LMS03}{\rm\cite{LMS03}}
Let $\bs{U}$ be a primitive lattice substitution system with expansive map $Q$
for the lattice $\varGamma = \bigcup_{i \le m} U^{}_i$  in $\RR^d$  and
suppose that every $\bs{U}$-cluster is legal. Let $\varGamma' =
\varGamma^{}_1+\cdots+\varGamma^{}_m$, where $\varGamma^{}_i =
\langle \varLambda^{}_i-\varLambda^{}_i \rangle^{}_{\ZZ}$. The following
assertions are equivalent.
\begin{enumerate}
\item\label{thm:item1:multisets} $\bs{U}$ has pure point diffraction spectrum
  (meaning that each $U^{}_i$ has this property); 
\item\label{thm:item2:multisets} $\bs{U}$ has pure point dynamical spectrum
(meaning that each $U^{}_i$ has this property); 
\item $\operatorname{dens}(\bs{U} \triangle (Q_{}^n \alpha+\bs{U}))
    \stackrel{n \to \infty}{\longrightarrow} \bs{0} = (0)^{}_{i\le m}$
for all $\alpha \in \varGamma'$;   
\item  A modular coincidence relative to $Q_{}^M \varGamma'$ occurs in
  $\varPhi_{}^M$ for some $M$; 
\item Each $U^{}_i$ is a regular model set,  $i \le m$, for the cut and project
  scheme 
\begin{equation*}
\begin{array}{ccccc}\RR^{d}_{} & \longleftarrow & \RR^{d}_{} \times
  \overline{\varGamma} & \longrightarrow & \overline{\varGamma} \\ 
&& \cup && \\ 
\varGamma &\longleftarrow & \tilde{\varGamma}  && 
\end{array}
\end{equation*}
\end{enumerate}
\vspace*{-5ex}\qed
\end{theorem}

Here, the \textit{pure point dynamical spectrum} is defined as follows: For a
dynamical system $(X,T)$ (see Eq.~\ref{eq:dynamical_system}) that has a
(unique) invariant probability measure $\mu$ associated to it (which is the
case for dynamical systems which arise from primitive substitutions), we have
the Hilbert space $L_{}^{2}(X,\mu)$ and the unitary operator
$B\!\!\!: L_{}^{2}(X,\mu) \to L_{}^{2}(X,\mu), \; f
\mapsto f \circ T$. If the eigenfunctions of $B$ span
$L_{}^{2}(X,\mu)$, then we have a pure point dynamical
spectrum. Note that the equivalence of~\ref{thm:item1:multisets}
and~\ref{thm:item2:multisets} can also be established in a more general
setting, see~\cite{BL03}.  

The model set in this theorem is with respect to a very particular cut and
project scheme. The key point is the internal group $\overline{\varGamma}$
which we now briefly explain. $\overline{\varGamma}$ is the $Q$-adic completion
(in terms of a \textit{profinite group}, see~\cite{Wil98,RZ00})
\begin{equation*}
\overline{\varGamma} = (\overline{\varGamma})^{}_Q = \lim_{\leftarrow k}
 \varGamma/Q_{}^k\varGamma' = \lim_{\leftarrow k} (\varGamma/\varGamma'
 \leftarrow \varGamma/Q\varGamma' \leftarrow \ldots \leftarrow
 \varGamma/Q_{}^k \varGamma' \leftarrow \ldots ) 
\end{equation*}
of $\varGamma$, supplied with the usual topology of a profinite group (which
makes it compact). We note that $\varGamma$ embeds naturally into
$\overline{\varGamma}$. Then, $\tilde{\varGamma}$ is the group $\{(t,t) \in
\RR^d \times \overline{\varGamma}  \mathbin|  t \in \varGamma \}$.

This gives a satisfactory approach to those systems which \textsl{are} model
sets, including an algorithm to test it. The latter, however, is of limited
value to disprove the model set property, unless one can see that no
coincidence can ever occur. For this situation, Frettl\"{o}h has recently
proved several sufficient criteria. Although they are not exhaustive, they are
easy to check and seem to cover many cases of relevance, see~\cite{F02} for
details. 

\section{Paperfolding sequence as model set \label{sec:paperfolding}}

To demonstrate the usefulness of the general setting, let us consider an
explicit example with a $p$-adic internal space.

The so-called regular paperfolding sequence\footnote
{\label{footnote:paper}
More generally, we can define a paperfolding sequence recursively:
the sequence $\{a_1, a_2,  \dots \}$ is called a paperfolding 
sequence iff $a_1 = -a_3 = a_5 = -a_7 = \dots$, and the remaining sequence 
$\{a_2, a_4, \dots \}$ is a paper folding sequence. We obtain the
regular paperfolding sequence for $a_n\in\{1,-1\}$ and writing $0$ for
$-1$.
}
starts as 
\begin{equation}\label{eq:paper_seq}
11011001110010011101100011001001\ldots
\end{equation}
and can be obtained by folding a sheet of paper repeatedly to the left,
see~\cite{DMFP82}: 
\begin{vchfigure}[ht]
\centerline{\includegraphics[scale=0.4]{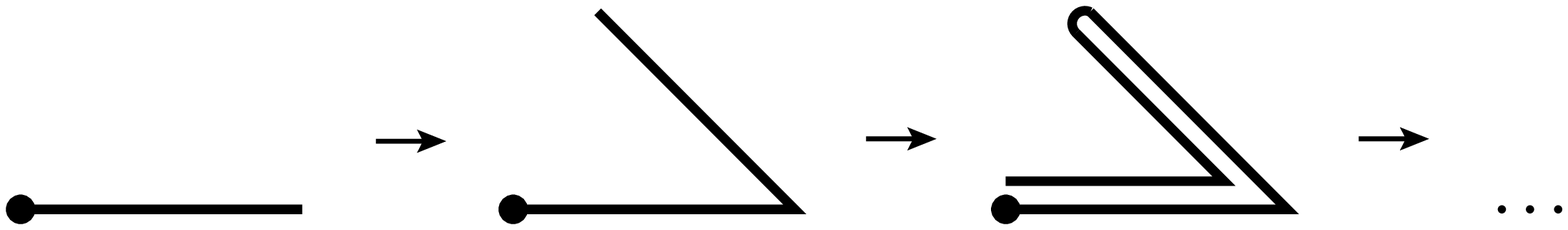}}
\end{vchfigure}

The sequence in Eq.~\ref{eq:paper_seq} is obtained by unfolding the (infinite)
stack and encoding a left (right) bend by $1$ ($0$).

An alternative description employs two steps: The first determines the unique
one-sided fixed point of the primitive $4$-letter substitution of constant
length\footnote
{
This is a primitive substitution of height $1$ which has a coincidence (after
two substitutions). Therefore it is pure point diffractive by a criterion of
Dekking, see~\cite{Dek78}.
}:
\begin{equation*}
\sigma: \begin{array}{lcl}
a & \to & ab \\ b & \to & cb \\ c & \to & ad \\ d & \to & cd, 
\end{array}
\end{equation*}
which starts as $abcbadcbabcdadcb\ldots$. The following second step maps $a$
and $b$ to $1$ and $c$ and $d$ to $0$, giving the
sequence in Eq.~\ref{eq:paper_seq}. 

Here, we change the point of view slightly, in that we consider two-sided (or
bi-infinite) fixed points of $\sigma$, of which there are precisely
two\footnote
{
The dynamical zeta function can be calculated with the method of Anderson and
Putnam~\cite{AP98}, and gives 
\begin{equation*}
\zeta(z) = \frac1{1-2\,z}.
\end{equation*}
From here, one sees that the number of fixed points of $\sigma^{m}_{}$ ($m \ge
1$) is $2^{m}$, which shows that $w^{}_1$ and $w^{}_2$ are the only
solutions of $\sigma(w)=w$ with $w$ a bi-infinite word.
}:
\begin{equation*}
\begin{array}{ccccccccc}
b|a & \to & cb|ab & \to & adcb|abcb & \to & \ldots & \to & w^{}_1 \\
d|a & \to & cd|ab & \to & adcd|abcb & \to & \ldots & \to & w^{}_2 
\end{array}
\end{equation*}
where $|$ denotes the seamline. Note that $w^{}_1$ and $w^{}_2$ differ only in
the first position to the left of the seamline, otherwise they are equal.

Let us represent the letters with intervals of equal length $1$, and points on
their left end, of type $a$, $b$, $c$ and $d$. Let $\varOmega^{}_{a}$ denote
the $a$-points, etc. Then, the substitution together with the fixed point
equation $\sigma(w) = w$ lead to the following system of equations:
\begin{equation*}
\begin{array}{lclcl}
\varOmega^{}_{a} & = & 2\,\varOmega^{}_{a} & \cup & 2\,\varOmega^{}_{c} \\
\varOmega^{}_{b} & = & (2\,\varOmega^{}_{a} + 1) & \cup & (2\,\varOmega^{}_{b}
+1) \\
\varOmega^{}_{c} & = & 2\,\varOmega^{}_{b} & \cup & 2\,\varOmega^{}_{d} \\
\varOmega^{}_{d} & = & (2\,\varOmega^{}_{c}+1) & \cup & (2\,\varOmega^{}_{d} +
1) \\ 
\end{array}
\end{equation*}
where also $\varOmega^{}_{a} \cup \varOmega^{}_{b} \cup \varOmega^{}_{c} \cup
\varOmega^{}_{d} = \ZZ$ by construction. With this, one quickly checks
that the first and the third equation lead to the unique solution
\begin{equation*}
\varOmega^{}_{a} = 4\, \ZZ, \qquad \varOmega^{}_{c} = 4\, \ZZ + 2
\end{equation*}
which reduces the other equations to
\begin{equation*}
\begin{array}{lclcl}
\varOmega^{}_{b} & = & (8\,\ZZ + 1) & \cup & (2\,\varOmega^{}_{b} +1 ) \\
\varOmega^{}_{d} & = & (8\,\ZZ + 5) & \cup & (2\,\varOmega^{}_{d} + 1). 
\end{array}
\end{equation*}
Since $\varOmega^{}_{b}$ and $\varOmega^{}_{d}$ are subsets of $\ZZ$, the
general solution is
\begin{equation*}
\begin{array}{lcl}
\varOmega^{}_{b} & = & \bigcup\limits_{m \ge 1} 2^{m+2}\,\ZZ + 2^{m} - 1 \\
\varOmega^{}_{d} & = & \bigcup\limits_{m \ge 1} 2^{m+2}\,\ZZ + 3 \cdot 2^{m} -
1,  
\end{array}
\end{equation*}
where the only remaining freedom consists in adding the singleton set $\{-1\}$
to either of them. This reflects the difference between the two fixed points,
$w^{}_1$ ($\{-1\}$ goes to $\varOmega^{}_{b}$) and $w^{}_2$ ($\{-1\}$ goes to
$\varOmega^{}_{d}$). 

If one now follows the construction of a canonical cut and project scheme as
derived in~\cite{BM02}, one finds that the autocorrelation topology is the
$2$-adic topology, and this completes $\ZZ$ (the set of differences) to
$\overline{\ZZ}^{}_2$, the compact group of $2$-adic integers. So we have
the cut and project scheme
\begin{equation*}
\begin{array}{ccccc}
\RR & \longleftarrow & \RR \times \overline{\ZZ}^{}_2 & \longrightarrow &
\overline{\ZZ}^{}_2 \\ && \cup && \\ && \tilde{L} && 
\end{array}
\end{equation*}
where $\tilde{L} = \{ (m,m) \mathbin| m\in \ZZ \}$ is a lattice. Our points
are now model sets in this scheme. Defining the windows (as subsets of
$\overline{\ZZ}^{}_2$) 
\begin{equation*}
W^{}_{a} = \overline{\varOmega^{}_{a}}, \quad W^{}_{b} =
\overline{\varOmega^{}_{b}}, \quad W^{}_{c} = \overline{\varOmega^{}_{c}},
\quad W^{}_{d} = \overline{\varOmega^{}_{d}},
\end{equation*}
one finds $W^{}_{a} \cap W^{}_{b} = \{-1\}$ and thus 
\begin{equation*}
\varOmega^{}_{a} = \varLambda(W^{}_{a}), \quad \varOmega^{}_{b} =
\varLambda(W^{}_{b}), \quad \varOmega^{}_{c} = \varLambda(W^{}_{c}), \quad
\varOmega^{}_{d} = \varLambda(W^{}_{d} \setminus \{-1\})
\end{equation*} 
for $w^{}_1$, while $\{-1\}$ moves from $\varOmega^{}_{b}$ to
$\varOmega^{}_{d}$ for $w^{}_2$.

As a regular model set, the paperfolding sequence is pure point
diffractive\footnote
{
It is an example of a limit periodic system.
}. 
More precisely, if $\omega = A\cdot \delta^{}_{\varOmega^{}_{a}} +
B\cdot\delta^{}_{\varOmega^{}_{b}} + C\cdot\delta^{}_{\varOmega^{}_{c}} + 
D\cdot\delta^{}_{\varOmega^{}_{d}}$, the diffraction measure reads
\begin{equation}\label{eq:paper_spectrum}
\begin{split}
\hat{\gamma}^{}_{\omega} = & \left| \frac{A+B+C+D}{4}\right|^2
\delta^{}_{\ZZ} \\ & + \sum_{m \text{ odd}} \left[ \left|
    \frac{A-B+C-D}{4}\right|^2 \delta^{}_{\frac{m}{2}} + \left|
    \frac{A-C}{4} \right|^2 \delta^{}_{\frac{m}{4}} + \sum_{r \ge 3} \left|
    \frac{B-D}{2^{r}} \right|^2 \delta^{}_{\frac{m}{2^{r}}} \right]
\end{split} 
\end{equation}

Finally, let us consider the binary reduction. One gets
\begin{equation*}
\begin{array}{lcl}
\varOmega^{}_{a} \cup \varOmega^{}_{b} & = & \bigcup\limits_{m \ge 0}
2^{m+2}\, \ZZ + 2^{m} - 1 \\
\varOmega^{}_{c} \cup \varOmega^{}_{d} & = & \bigcup\limits_{m \ge 0}
2^{m+2}\, \ZZ + 3 \cdot 2^{m} - 1
\end{array}
\end{equation*}
plus $\{-1\}$ added to one of them. Clearly, these are again regular $2$-adic
model sets, and thus also pure point diffractive. To summarise:

\begin{theorem}
The quaternary regular paperfolding sequences $w^{}_1$ and $w^{}_2$ are
regular $2$-adic model sets, with pure point diffraction spectrum as given
in Eq.~{\rm\ref{eq:paper_spectrum}}. Also, the binary reduction is a regular
$2$-adic model set, hence also pure point diffractive. \qed
\end{theorem}

This structure is then inherited by the entire LI-class\footnote
{
Two structures $\varLambda^{}_1$ and $\varLambda^{}_2$ are \textit{locally
indistinguishable}\index{locally indistinguishable}\index{locally
isomorphic|see{locally indistinguishable}}\index{LI|see{locally
  indistinguishable}} (or \textit{locally isomorphic} or \textit{LI}) if each
patch of $\varLambda^{}_1$ (essentially, the intersection of
$\varLambda^{}_{1}$ with a compact set) is, up to translation, also a patch
of $\varLambda^{}_2$ and vice versa. The corresponding equivalence class is
called \textit{LI-class}.
}
of the paperfolding sequence. The members can be obtained via different
folding sequences\footnote
{
We get the paperfolding in Eq.~\ref{eq:paper_seq} by 
setting $1 = a^{}_{1} = a^{}_{2} = a^{}_{4} = \ldots = a^{}_{2^{n}} =
\ldots$ (see Footnote~\ref{footnote:paper}), because we only fold in one
direction. Using a different folding sequence, which corresponds to the
$a^{}_{2^n}$ not being equal, we get a different paperfolding sequence, but
this one is LI to the one in Eq.~\ref{eq:paper_seq} and vice versa and
therefore in the same LI-class. So, all such paperfolding sequences have the
same diffraction spectrum, namely Eq.~\ref{eq:paper_spectrum} in the binary
reduction, i.e., $A=B$ and $C=D$.
}, 
see~\cite{DMFP82,MF82} and references therein for details. Further examples
along similar lines can be found in \cite{BM02,BMS98,S02}. 

\section{Systems with disorder\label{sec:disorder}}

Here, we consider diffraction properties of stochastic point sets. Simple,
well-understood examples are Bernoulli subsets of lattices or model
sets~\cite{B00,BM98}, and certain lattice gases, which can be analysed using
elementary methods from stochastics. This approach has recently been
generalised considerably~\cite{K01} to cover stochastic selections from rather
general Delone sets. The results prove the folklore claim that uncorrelated
random removal of scatterers has two effects, namely reducing the overall
intensity of the diffraction of the fully occupied set, without changing the
relative intensities, and adding a white noise type constant diffuse
background. The influence of disorder due to thermal fluctuations is
discussed in~\cite{Hof95b}.

A prominent class of stochastic point sets are random
tilings~\cite{Hen99,RHHB98,R99}. Diffraction properties of these tilings are 
understood for systems without interaction, for one-dimensional Markov
systems, and for product tilings~\cite{HB00,BH00,H01}. Systems with
interaction are generally difficult to analyse. Here, only few results are
available for certain exactly solvable models from statistical mechanics with
crystallographic symmetries, whose autocorrelation can be computed
explicitly~\cite{HB00,BH00,H01,H99}. Symmetries of a stochastic point set are
understood to be symmetries on average. For a detailed discussion of this
concept, see~\cite{RHHB98,BH00}. 

Most examples of random tilings with quasicrystallographic symmetries are
obtained from ideal quasicrystallographic tilings by relaxing the allowed
local configurations. Since, as for ideal tilings, random tiling
coordinates may be lifted into internal space via the $\star$-map,
the random tiling ensemble possesses a so-called \textit{height
  representation}\index{height representation}  
which, in a way, can be understood as a description on the basis of a
deviation from a model set. At present, there are no rigorous results
concerning diffraction properties of random tilings with height representation
in dimension $d\ge 2$. Henley~\cite{Hen99} argues, using elasticity theory for
the free energy of such an ensemble, that the discrete part of the diffraction
spectrum  only consists in the trivial Bragg peak at the origin in dimension
$d\le 2$, since the width of the distribution of scatterer positions in
internal  space diverges with the system size. In dimensions $d>2$, the
distribution width converges with the system size, implying a non-trivial
discrete part in the diffraction spectrum. Even less is known about the nature
of the continuous part in $d>1$, though absolute continuity is expected. 
For a numerical investigation of diffraction properties of the randomized
Ammann-Beenker tiling, see~\cite{H01}.
In what follows, we will focus on stochastic disorder in random tilings, 
mainly in the one-dimensional case, because the understanding of the higher
dimensional situation is still rather incomplete.

The diffraction of 1D random tilings\footnote
{
A \textit{(1D binary) random tiling}\index{random tiling}~\cite{Hen99,RHHB98} 
is a covering of the real line with two intervals of fixed lengths $u$
and $v$ without gaps or overlaps. Associated with each random tiling is the set
$\varLambda$ of left endpoint  positions of its intervals. 
We call two random tilings \textit{equivalent} if they are equal 
up to a translation. For each equivalence class, we choose a representative
with $0\in\varLambda$. The \textit{random tiling ensemble} is the set of all
non-equivalent random tilings.
}
has been investigated previously~\cite{BH00}. 1D binary random tilings have 
a non-trivial pure point part iff they have a rational interval length ratio 
$\alpha=u/v$.

\begin{theorem}\label{thm:1Dbinary_random}{\rm\cite{BH00}}
The natural density of $\varLambda$ exists with probabilistic certainty and is
given by $d=(pu+qv)_{}^{-1}$. If $\omega=\delta^{}_{\varLambda}=\sum_{x\in
  \varLambda}\delta^{}_x$ denotes the corresponding stochastic Dirac comb, the
autocorrelation $\gamma_\omega$ of $\omega$ also exists with probabilistic
certainty and is a positive definite pure point measure. 
The diffraction spectrum consists, with probabilistic certainty, of a pure
point (Bragg) part and an absolutely continuous part, so  
$\hat{\gamma}^{}_{\omega} = (\hat{\gamma}^{}_{\omega})^{}_{\textnormal{pp}} +
(\hat{\gamma}^{}_{\omega})^{}_{\textnormal{ac}}$.  

If $\alpha=u/v$, the pure point part is
\begin{equation*}
(\hat{\gamma}^{}_{\omega})^{}_{\textnormal{pp}} = d_{}^2 \cdot \left\{ 
\begin{array}{ll}
\delta^{}_0 & \mbox{if } \alpha \notin \QQ \\
\sum_{k \in (1/\xi) \ZZ} \delta_k & \mbox{if } \alpha \in \QQ
\end{array}
\right.
\end{equation*}
where, if $\alpha \in \QQ$\,, we set $\alpha=a/b$ with coprime $a,b \in \ZZ$
and define $\xi=u/a=v/b$. 
The absolutely continuous part
$(\hat{\gamma}^{}_{\omega})^{}_{\textnormal{ac}}$ can be represented by the
continuous function 
\begin{equation*}
g(k) = 
\frac{d\cdot pq \, \sin_{}^2 (\pi k\, (u-v))}
{p \, \sin_{}^2(\pi k \, u)+ q \, \sin_{}^2 (\pi k \, v)- pq \, \sin_{}^2 (\pi
  k\,  (u-v))}
\end{equation*}
which is well defined for $k\, (u-v)\notin \ZZ$.
It has a smooth continuation to the excluded points.
If $\alpha$ is irrational, this is $g(k)=0$ for $k(u-v)\in\ZZ$ with $k\neq 0$
and 
\begin{equation*}
g(0)=\frac{d\cdot pq\, (u-v)_{}^2}{p\, u_{}^2+ q\, v_{}^2- pq\, (u-v)_{}^2} =
d\,\frac{pq\, (u-v)_{}^2}{(p\, u+q\, v)_{}^2}
\end{equation*}
For $\alpha=a/b\in \QQ$ as above, it is $g(k)=0$ for $k\, (u-v)\in\ZZ$, 
but $k\, u\notin\ZZ$ (or, equivalently, $k\, v\notin\ZZ$), and
\begin{equation*}
g(k) = d\,\frac{pq\, (a-b)_{}^2}{(p\, a+q\, b)_{}^2}
\end{equation*}
for the case that also $k\, u\in\ZZ$. \qed
\end{theorem}

The most prominent 1D random tiling is the Fibonacci random tiling\footnote
{
A \textit{Fibonacci random tiling} is a random tiling with interval lengths 
$u=\tau=(1+\sqrt{5})/2$ and $v=1$, with occupation probabilities 
$p=1/\tau$ and $q=1-p=1/\tau^2$ of the intervals (almost surely).
Each interval endpoint of a representative of a Fibonacci random tiling
belongs to the module $\ZZ[\tau]=\{m\tau+n \mathbin| m,n\in\ZZ\}$.
Every (ideal) Fibonacci tiling also appears as a Fibonacci random tiling.
}.
According to be above theorem, its diffraction spectrum consists 
with probabilistic certainty of a trivial Bragg peak at the origin and of an 
absolutely continuous background, see Fig.~\ref{fig:background}.
The absolutely continuous background shows localised, bell-shaped needles of
increasing height at sequences of points scaling with the golden ratio $\tau$.
This is reminiscent of the perfect Fibonacci tiling.
\begin{vchfigure}[tbp]
\centerline{\includegraphics[scale=0.8]{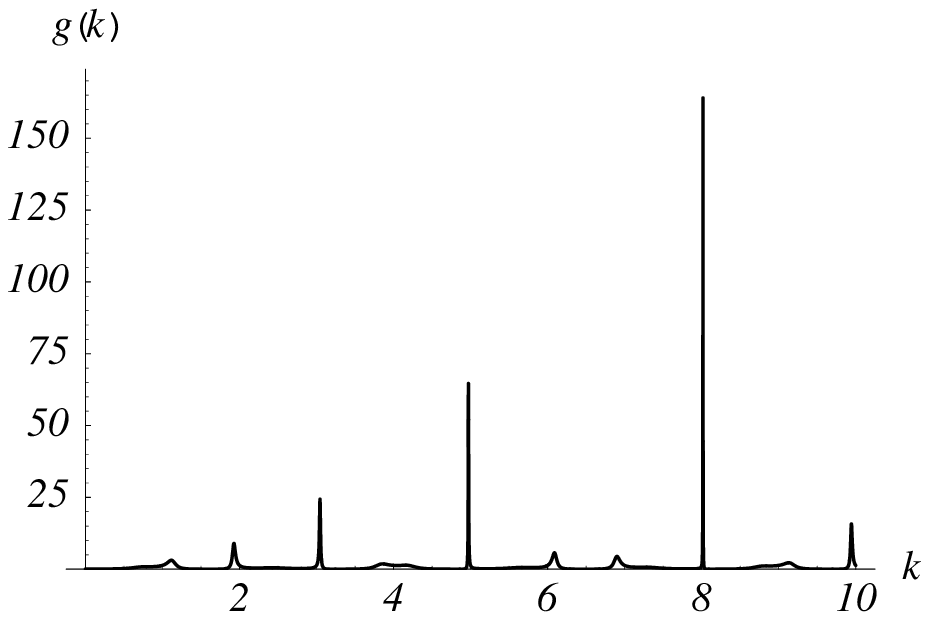}}
\vchcaption{Absolutely continuous background of a Fibonacci random tiling.}
\label{fig:background}
\end{vchfigure}

In dimensions $d\ge2$, properties of the autocorrelation are known only
for certain simple systems of statistical mechanics with crystallographic
symmetries which can be interpreted in terms of dimer systems,
see~\cite{BH00,H99,BHHR98}. 
This includes the domino and the lozenge tiling, the Ising lattice gas, and
others. Here, the asymptotic behaviour of the autocorrelation coefficients can
be computed explicitly, leading to proofs of existence of an absolutely
continuous part in addition to a pure point part.

\begin{theorem}\label{thm:Ising_gas}{\rm\cite{BH00}}
Away from the critical point, the diffraction spectrum of the Ising lattice gas
almost surely exists, is $\mathbb Z^2$-periodic and consists of a pure point
and an absolutely continuous part with continuous density. The pure point part
reads 
\begin{eqnarray*}
&1.& T>T_c: (\widehat{\gamma}_\omega)_{pp}=\frac{1}{4}\sum_{{\bs{k}}\in\mathbb
  Z^2} \delta_{\bs{k}}\\ 
&2.& T<T_c: (\widehat{\gamma}_\omega)_{pp}=\rho^2\sum_{{\bs{k}}\in\mathbb Z^2}
  \delta_{\bs{k}}, 
\end{eqnarray*}
where the density $\rho$ is the ensemble average of the number of scatterers
per unit volume. At the critical point, the diffuse scattering diverges when
approaching the lattice positions of the Bragg peaks.
\end{theorem}

We now analyse the effect of a cut and project setup on diffraction properties.
For 1D random tilings, an embedding into $\mathbb R^2$ is given as follows.
Each position $x\in\varLambda$ of an element of the random tiling ensemble 
may be written in the form $x=mu+nv$, where $m,n\in\ZZ$.
If $\alpha=u/v\not\in \QQ$, the numbers $m,n$ are uniquely determined.
If $\alpha\in\QQ$, uniqueness is achieved by parametrising $x=0$ by $m=n=0$ and
incrementing (decrementing) $m$ by addition of a $u$-interval to the right
(left), and likewise with $n$ for $v$-intervals.
Identifying the unique coordinates $m,n$ with points $(m,n)\in\ZZ_{}^2$, we map
a random tiling to a bi-infinite, directed walk on the edges of the square
lattice. We call two walks equivalent if they are equal up to a translation.
For each equivalence class of walks, we may choose a representative which 
passes through the origin in $\ZZ_{}^2$. This establishes a one-to-one
correspondence between non-equivalent random tilings and non-equivalent
bi-infinite, directed walks.

Let us restrict to Fibonacci random tilings.
Recall that the (ideal) Fibonacci tilings may be obtained
within the cut and project setup by (orthogonal) projection of lattice
points of a scaled copy of $\ZZ^2$ confined to a suitable strip onto the 
subspace with irrational slope $1/\tau$.
This way, we obtain a one-to-one correspondence between (random) tiling
coordinates in direct space and in internal space via the $\star$-map, given
by $(m\tau+n)^{\star}=m\tau'+n$, where $\tau'=-1/\tau$ is the algebraic
conjugate of $\tau$. The value $x^{\star}$ of a tiling coordinate $x$ is also
called its \textit{height}, and the above collection of direct and internal
space together with the canonical projections is also called \textit{height
  representation}\index{height representation} of the Fibonacci random tiling
ensemble. 

In the following, we will consider the distribution of scatterer positions of
Fibonacci random tilings in internal space.
We restrict ourselves to patches of Fibonacci random tilings 
of $M$ consecutive intervals on the positive half-axis, starting at $x=0$.
Following~\cite{H01}, we consider the occupation probability for the position 
of the rightmost interval. Since the random tiling patch is a Bernoulli
system, the probability of the position being $x=m\tau+(M-m)$, or
equivalently $x_{}^{\star} =m\tau' + (M-m)$ in internal space, is given by
\begin{equation*}
\tilde{\rho}(M, x_{}^{\star}) = \binom{M}{m}\,  p_{}^{m} \,  q_{}^{M-m}.
\end{equation*}
According to the theorem of de Moivre-Laplace, the binomial distribution 
(with mean $\mu=Mp$ and variance $\sigma^2=Mpq$) may 
be approximated by the Gaussian distribution for large fixed $M$, yielding
\begin{equation*}
\rho(M, x_{}^{\star}) = \sqrt{\frac{1}{\pi} \, \frac{\tau} {2 M}}\;
\exp{ \left[-\frac{\tau}{2M}\, {x_{}^{\star}}_{}^2\right]}.
\end{equation*}
Note that in the limit $M\to \infty$, the admissible positions $x_{}^{\star}$
lie dense in internal space. We fixed the normalisation such that the integral
of the density over internal space equals unity. To obtain the distribution of
all interval positions of random tiling patches of length $N$, we sum over all
endpoint positions of patches with $n \le N$ intervals and normalise\footnote
{
A more natural normalisation may be the density of points instead of unity,
see below.
},
\begin{equation}\label{eq:internaldistr}
\rho(x_{}^{\star}) = \frac{1}{N} \, \sum_{n=1}^N \rho(n, x_{}^{\star})
\simeq \frac{1}{N} \, \int_{n=0}^N \rho(n, x_{}^{\star})\, {\rm d}n.
\end{equation}
In the second equation, we approximated the sum to leading order in $N$ using
the Euler-Maclaurin summation formula.
This leads to the following result.

\begin{theorem}\label{thm:intdistr}
The distribution $\rho(x_{}^{\star})$ of scatterer positions of
Fibonacci random tiling patches in internal space is, to leading order in the
patch size $N$, given by 
\begin{equation*}
\rho(x_{}^{\star})= \sqrt{\frac{\tau}{2N}}\; f\left(
\sqrt{\frac{\tau}{2N}}\cdot x_{}^{\star}\right), 
\qquad 
 f(z) = 2 \left( \frac{e_{}^{-z_{}^2}} {\sqrt{\pi}} - 
 |z| \, \operatorname{erfc}\left(|z|\right)\right),
\end{equation*}
where $\operatorname{erfc}(x)=\frac{2}{\sqrt{\pi}}\int_x^{\infty}
e_{}^{-t_{}^2} {\rm d}t$  denotes the complementary error function.
\qed
\end{theorem}

The function $f(z)$ is shown in Fig.~\ref{fig:fibo}. 
\begin{vchfigure}[tbp]
\centerline{\includegraphics[scale=0.8]{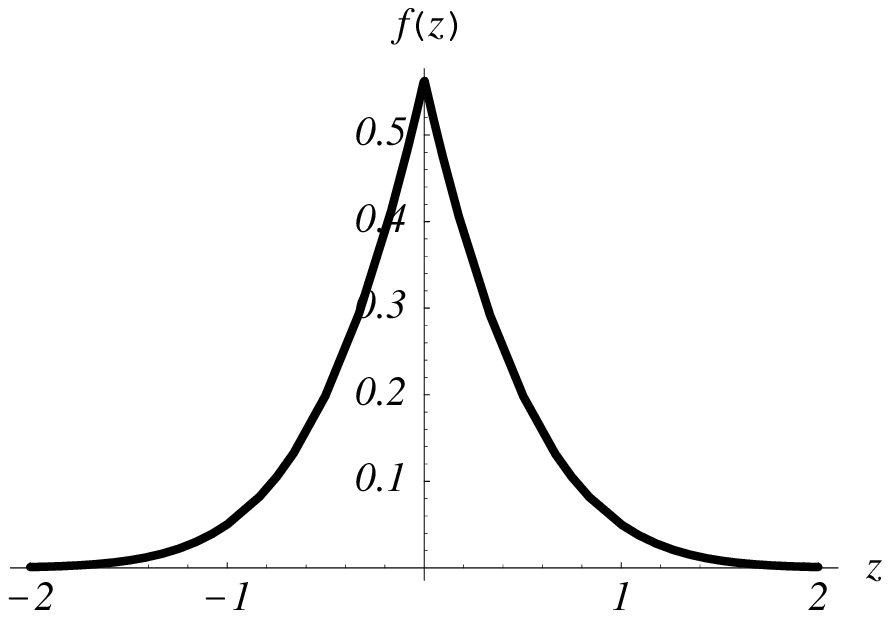}}
\vchcaption{Distribution of scatterer positions in internal space
for Fibonacci random tilings.}
\label{fig:fibo}
\end{vchfigure}
Note that the width of the distribution grows with the system size as 
$\sqrt{N}$, but the distribution itself is not a Gaussian, as usually
believed\footnote 
{
For a comparison with numerical simulations of the Fibonacci
random tiling ensemble and for the example of the two-dimensional
Ammann-Beenker random tiling, see~\cite{H01}.
}.

This result was derived for patches of $N$ intervals, starting from the
origin in positive direction.
Since the result is symmetric in $x_{}^{\star}$, it is also valid for tiling
patches with $N$ intervals, starting from the origin in negative direction,
hence also for tiling patches with $2N$ intervals with $N$ intervals on 
the negative and on the positive axis each.

Within the cut and project scheme, the Dirac comb of a model set may be
characterised as a sum of point scatterers over the projected lattice points
\begin{equation*}
\omega = \sum_{x\in L} 1^{}_{\Omega}(x_{}^{\star})\, \delta^{}_{x},
\end{equation*}
weighted by the characteristic function of the (compact) window in 
internal space. For diffraction of quasicrystallographic random tilings, the
common approach~\cite{Hen99} is to investigate properties of an averaged 
structure given by the Dirac comb weighted by the occurrence probability of
each scatterer within the random tiling ensemble. The averaged distribution in
internal space need not exist for the infinite tiling, as we showed in the
previous section. It is generally believed~\cite{Hen99} that it exists in
dimensions $d>2$, the case $d=2$ being marginal with a logarithmic growth of
the  distribution width with the system size, leading to the statement that
there is no non-trivial discrete component in the diffraction spectrum for
$d\le 2$ (for aperiodic systems). 

As argued in Eq.~\ref{eq:internaldistr}, the averaged structure
may be written in the form
\begin{equation}\label{eq:rtcomb}
\omega = \sum_{x\in L} \varphi(x_{}^{\star})\, \delta^{}_{x},
\end{equation}
where the support of $\varphi$ is generally the whole internal space.
Note that this object is generally ill-defined, since summation is over 
a dense set. In averaging, one loses information about correlations between
scatterers, so that the analysis will at most yield information about
the discrete part of the diffraction spectrum, but not about the
continuous part, see the theorem below. 
We investigate under which conditions Eq.~\ref{eq:rtcomb} 
defines a tempered  distribution. To this end, we assume that $\varphi$
vanishes sufficiently rapidly at infinity, which includes the Gaussian (and,
in a certain sense, characteristic
functions). Following~\cite{H01}, we consider the special
situation of a Euclidean internal space $H=\RR^m$ and assume that the
canonical projections $\pi$ and $\pi_{\textnormal{int}}$  are both dense and
one-to-one. We denote by $\operatorname{vol}(FD)$ the volume of a fundamental
domain of the lattice $\tilde L\in \RR^d\times H$ w.r.t. the product measure
of the Lebesgue measures on $\RR^d$ and on $H$.
We denote the dual lattice of $\tilde{L}$ by 
$(\tilde{L})^* = \{ \tilde{x} \in \RR^d\times H \mathbin| \langle \tilde{x},
\tilde{y}\rangle \in  \mathbb Z \mbox{ for all } \tilde{y} \in \tilde{L} \}$,
and its projection by $L_{}^{*}=\pi((\tilde{L})_{}^{*})$.

\begin{theorem}\label{thm:R02}{\rm\cite{R02}} 
Assume $\varphi\!: \RR^m \to \CC$ continuous and  
$\lim_{y\to\infty} |y|_{}^{m+1+\alpha}\, \varphi(y)=0$ for some $\alpha>0$.
Then, the weighted Dirac comb in Eq.~{\rm\ref{eq:rtcomb}} is a translation
bounded measure. It has the unique autocorrelation 
\begin{equation*}
\gamma^{}_{\omega}=\sum_{z\in L} \eta(z)\, \delta^{}_{z}, \qquad 
\eta(z)=\frac{1}{\operatorname{vol}(FD)}\,
\int_{\RR^m}\varphi(u)\,\overline{\varphi(u-z_{}^{\star})}\,{\rm d}u, 
\end{equation*}
being a translation bounded, positive definite pure point measure.
Its Fourier transform is a positive pure point measure.
If $\pi$ and $\pi_{int}$ are orthogonal projections, it is explicitly given by
\begin{equation*}
\hat{\gamma}^{}_{\omega}=\frac{1}{\operatorname{vol}(FD)^2}\sum_{y\in L_{}^{*}}
|\hat{\varphi}(-y_{}^{\star})|_{}^2\, \delta^{}_{y}, 
\end{equation*}
where $\hat{\varphi}$ denotes the Fourier transform of $\varphi$. \qed
\end{theorem}
A natural choice for the normalisation of the function $\varphi$ arises
form the observation that for Dirac combs satisfying the assumptions of 
the above theorem the density of points $\rho$ exists,
\begin{equation}
\rho = \lim_{n\to\infty} \frac{1}{\operatorname{vol}(B_n)}\,\omega(B_n) = 
\frac{1}{\operatorname{vol}(FD)} \int_{\RR^m} \varphi (u) \, {\rm d}u.
\end{equation}

The above theorem sheds light onto the example of the Fibonacci random tiling.
Here, the internal distribution may be described by a sequence 
of distributions of increasing width but constant mass.
The limit of the corresponding sequence of measures will have a trivial
discrete part and a continuous part, whose form may be compared to the 
above results.

\section{Outlook\label{sec:outlook}}

For our discussion of pure pointedness, we made substantial use of the uniform
discreteness of $\varDelta^{\textnormal{ess}}_{}$. This can certainly be
relaxed, as Theorem~\ref{thm:pp_almostper} shows, but things become
considerably more involved beyond this ``barrier''. This is also intimately
related to stepping into the territory of mixed spectra, which seems
particularly timely.

Pure pointedness of the diffraction is equivalent to strong almost periodicity
of the autocorrelation. More generally, one can show that any autocorrelation
$\gamma^{}_{\omega}$ possesses a unique decomposition into a strongly almost
periodic part and a weakly almost periodic part with zero volume average,
compare~\cite{LA90}. So, we get 
\begin{equation*}
\gamma^{}_{\omega} = (\gamma^{}_{\omega})^{}_{\textnormal{sap}} +
(\gamma^{}_{\omega})^{}_{0\textnormal{--wap}} 
\end{equation*}
where ``weak'' refers to the weak topology in relation to the strong (product)
topology. 

The Fourier transform of $(\gamma^{}_{\omega})^{}_{\textnormal{sap}}$ is a
pure point measure, while that of
$(\gamma^{}_{\omega})^{}_{0\textnormal{--wap}}$ is continuous~\cite{LA90}, so
that one has full control of this question on the level of the
autocorrelation. Also, important issues of the diffraction of random tilings
can be formulated and understood in this context, but most
results are folklore, and still need to be proved. Furthermore, there is no 
such decomposition known that would allow a distinction of absolute versus 
singular continuity. It is highly desirable to improve this situation in the 
future.

Finally, even if all the spectral questions were settled, the big remaining
question is how to characterise the homometry class, i.e., the class of
measures with a given autocorrelation (and hence diffraction). This is part of
the inverse problem, where results are rare at present. So, the question of
the title should now be replaced by another one:
\begin{center}
``Which distributions of matter are homometric?''
\end{center}

\section*{Acknowledgements}

A considerable amount of the work summarised here was done during the six
years of the DFG focus program on quasicrystals (project number: Ba 1070/7),
and in constant interaction with several colleagues, both within and outside
this program. The financial support for mutual visits and a series of
workshops is gratefully acknowledged. Robert Moody acknowledges the support of
an NSERC Discovery Grant in this work. Christoph Richard acknowledges funding
by the DFG. Michael Baake and Christoph Richard are
grateful to the Erwin Schr\"{o}dinger International Institute for
Mathematical Physics in Vienna for support during a stay in December 2002,
where the manuscript was completed. 
We also thank Moritz H\"offe for important discussions.

\end{document}